\documentclass[a4paper,10pt,twocolumn]{revtex4}
\usepackage[english]{babel}
\usepackage{indentfirst}
\usepackage{fancyhdr}
\usepackage{graphicx}
\usepackage{amsmath}
\usepackage{amssymb}
\usepackage{amsthm}
\usepackage{mathrsfs}
\usepackage{lscape}
\usepackage{bm}
\usepackage{subfigure}
\usepackage{caption}
\usepackage{float}
\usepackage{longtable}
\usepackage{nicefrac}
\usepackage{graphics}
\usepackage{color}

\begin{document}
\title{Single photon absorption by a single quantum emitter}
\author{D. Pinotsi and A. Imamo\~glu}
\affiliation{Institute of Quantum Electronics, ETH Zurich, 8093
Zurich, Switzerland}

\vspace{-3.5cm}

\date{\today}
\begin{abstract}
We show that a three-level lambda quantum emitter with equal
spontaneous emission rates on both optically active transitions can
absorb an incident single photon pulse with a probability approaching unity,
provided that the focused light profile matches that of the emitter
dipole emission pattern. Even with realistic focusing geometries,
our results could find applications in long-distance
entanglement of spin qubits.
\end{abstract}
\maketitle

Optically thick systems such as an ensemble of atoms or a single atom  in a cavity play a key role in
quantum optics. Various applications of those schemes have been proposed, such as nonlinear optics using
slow light \cite{FIM05} and flying qubit to stationary qubit conversion \cite{Ciraczoller97prl}. It is
typically assumed that a single emitter in \textit{free} space will only have a small effect on incoming
radiation. Recent theoretical efforts aimed at understanding the fundamental limits on the strength of
this coupling \cite{vanenk+kimble01pra,vanenk04pra} as well as at developing novel methods for strong
light focusing \cite{leuchs00opt.comm.}. Experimental results already show that laser extinction of up
to 12\% is possible \cite{Vamivakas07,Sandoghdar107,Sandoghdar207}.

It is already established \cite{sheppard97} that matching the
\textit{dipole} emission profile of an emitter with a focused laser
is a necessary condition for achieving perfect absorption. However,
it was shown by van Enk \cite{vanenk04pra} that this is not
sufficient, since for a two-level emitter absorption is necessarily
followed by emission of a photon of identical first order coherence
properties; as a consequence, the detected transmitted or reflected
power is never zero.

We show here that if the emitter is a 3-level $\Lambda$ system,
then having equal spontaneous emission rates on both optical
transitions, along with matched dipole emission and focused laser
profiles is sufficient for \textit{perfect} extinction of light.
The scheme we describe does not require any additional coherent
coupling and in this way it is distinguished from EIT type systems
\cite{harris90prl}. In particular, we consider the interesting
limit of \textit{single photon} absorption as well as
possibilities of intra-conversion of a flying qubit to a
stationary qubit and entanglement of distant qubits. We assume
that the lower states of the 3 level $\Lambda$ system have a long
coherence time; this is the case for a single trapped atom/ion, or
a solid-state emitter provided that the two lower states are spin
states. To analyze single-photon absorption, we use the cascaded
quantum systems approach \cite{carmichael93prl}, since we have a
quantum system (target emitter) driven by another quantum
system (single photon). Using the source-field expression, we
express the incident photon annihilation/creation operators using
the time-retarded dipole operators of its source. We
further assume that this source S consists of a three level
quantum system whose
$\left|0\right\rangle\rightarrow\left|3\right\rangle$ transition
is driven resonantly via a classical laser pulse $\Omega_L(t)$.

\begin{figure}[hb]
\includegraphics[width=0.5\textwidth]{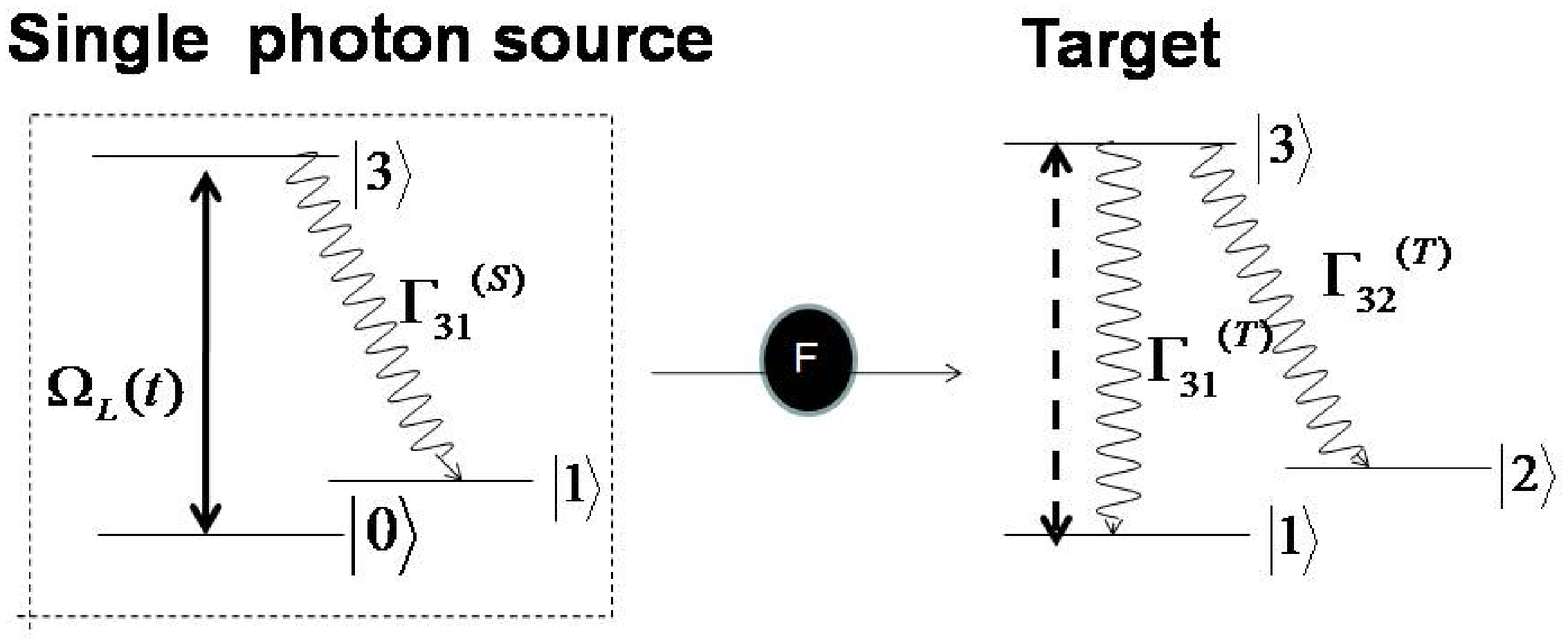}
\vspace{-3cm}
\caption {\small{Schematic of the cascaded system analysed in the text.
 The source atom on the left, generates a single-photon pulse upon
 excitation by a coherent laser field. The generated single photon
 is focused on the target atom to ensure maximum overlap with the target's atom
 dipole emission pattern. The Faraday
 rotator (F) ensures unidirectional coupling.}}
\end{figure}

The single-photon pulse emitted from the
$\left|3\right\rangle\rightarrow\left|1\right\rangle$ transition
($\omega_{31}^{(S)}$) of the source couples to the target atom and
drives its $\left|1\right\rangle\rightarrow\left|3\right\rangle$
transition ($\omega_{31}^{(T)}=\omega_{31}^{(S)}$). Following van
Enk \cite{vanenk04pra}, we introduce a parameter $\eta$ to
describe the overlap between the incoming single-photon field and
the dipole emission profile. In this formalism, the positive
frequency component of the electric field operator can be written
as the sum of two parts, one that corresponds to the relevant
dipole mode and a second for the remaining modes,
\begin{equation}
\widehat{E}_{tot}^{(+)}=\widehat{E}_{dip}^{(+)}+\widehat{E}_{nondip}^{(+)}
\end{equation}
If we consider an incoming coherent field such that $\left\langle
\widehat{E}_{dip}^{(+)}\right\rangle=\beta$ and $\left\langle
\widehat{E}_{nondip}^{(+)}\right\rangle=\alpha$ then $\eta$ is a
dimensionless number that relates the intensity of the field at
the target's position to the contribution of the dipole field:
\textbf{\begin{equation}
\eta=\frac{\left|\beta\right|^2}{\left|\alpha\right|^2+\left|\beta\right|^2}
\end{equation}}
$\eta = 1$ implies perfect overlap between the focused field and
the emission pattern of the emitter.

The cascaded quantum systems approach
\cite{carmichael93prl,gardiner93pra,gardiner85pra,kochan94pra} provides a description of the dissipative
dynamics of the two emitters that is equivalent to the master equation: here, a stochastic wave-function
describing the system evolves according to the Schr\"{o}dinger equation:
\begin{equation}
\frac{d}{dt}\left|{\psi_c}(t) \right\rangle=\frac{1}{\imath\hbar}H_{eff} \left|{\psi_c}(t)\right\rangle
\end{equation}
with the non-Hermitian Hamiltonian:
\begin{equation}
H_{eff}=H_S+H_T+H_{ST}-\frac{\imath\hbar}{2}\sum_{k}\hat C_k^{\dag}\hat C_k
\end{equation}

This coherent evolution of the stochastic wavefunction is interrupted by \textit{gedanken} measurement
events described by the collapse operators $\hat C_k$. In $H_{eff}$, $H_S,H_T$  denote the free source
and target Hamiltonian and $H_{ST}$ describes the interaction between them. The collapse operators are
in turn given by:
\begin{eqnarray}
\hat{C_1}&=&\sqrt{\Gamma_{31}^{(S)}}\sigma_{13}^{(S)}+\sqrt{\Gamma_{31}^{(T)}\eta}~\sigma_{13}^{(T)}\\
\hat{C_2}&=&\sqrt{\Gamma_{31}^{(T)}(1-\eta)}~ \sigma_{13}^{(T)}\\
\hat{C_3}&=&\sqrt{\Gamma_{32}^{(T)}}\sigma_{23}^{(T)}
\end{eqnarray}
By substituting the expressions for $\hat C_k$'s, we find \cite{carmichael93prl}:
\begin{eqnarray}
H_{eff}&=&\hbar\Omega_L(t)(\sigma_{03}^{(S)}+\sigma_{30}^{(S)})
-\imath\hbar\frac{\Gamma_{31}^{(S)}}{2}\sigma_{33}^{(S)}-\nonumber\\
&&-\imath\hbar\frac{\Gamma_{31}^{(T)}+\Gamma_{32}^{(T)}}{2}\sigma_{33}^{(T)}-\nonumber\\
&&-\imath\hbar\sqrt{\Gamma_{31}^{(S)}\Gamma_{31}^{(T)}\eta}~\sigma_{13}^{(S)}\sigma_{31}^{(T)}
\end{eqnarray}
where we assumed that the laser is resonant with the
$\left|0\right\rangle_S\rightarrow\left|3\right\rangle_S$
transition of the source atom, $\sigma_{ij}=\left|i\right\rangle\left\langle j\right|$ are the
projection ($i=j$) or the lowering/raising ($i\neq j$) operators
for the atomic states and $\Gamma_{ij}$ denote the spontaneous
emission rates of the source and target atoms.

The last term of the effective Hamiltonian $(8)$ is explicitly
unidirectional, since it is proportional to
$\sigma_{13}^{(S)}\sigma_{31}^{(T)}$ which corresponds to the
excitation of the target atom by the single photon generated by
the source. However, there is no term
$\sigma_{31}^{(S)}\sigma_{13}^{(T)}$ which would account for the
reverse process: the target atom absorbs photons emitted earlier in time by the source atom, 
while the backward scattered
photons from the target atom do not interact with the source.
Experimental implementation of this unidirectional feature of
coupling is achievable using a Faraday rotator placed in between
the source and the target emitters, as depicted in Figure~1.

Quantum jumps described by the collapse operators correspond to the detection of a photon with a
definite energy and emission profile. The collapse operator $\widehat{C}_1$ denotes the
\textit{gedanken} detection of a scattered photon in the incoming mode; such a detection event has
contributions from photons emitted in the $\left|3\right\rangle_S\rightarrow\left|1\right\rangle_S$ and
$\left|3\right\rangle_T\rightarrow\left|1\right\rangle_T$ transitions. However, a photon emitted in the
latter transition has a probability $\eta$ for coupling into the incoming mode. The collapse operator
$\widehat{C}_2$ accounts for a photon that is emitted by the target atom, that does not couple to the
incoming radiation mode. Finally, $\widehat{C}_3$ denotes the detection of a photon emitted in the
$\left|3\right\rangle_T\rightarrow\left|2\right\rangle_T$ transition.

Initially, both emitters (source and target) are assumed to be in their ground states;
$\left|\psi_{in}\right\rangle=\left|0_S,1_T\right\rangle$. The consequence of the desired
deterministic photon absorption process is the projection of the two-emitter wave-function to a state
where the source atom is in state $\left|1\right\rangle$ and the target atom is in state
$\left|2\right\rangle$ ($\left|a\right\rangle=\left|1_S,2_T\right\rangle$). By simulating the
dynamical evolution of the system wavefunction according to the effective non-Hermitian Hamiltonian,
interrupted at random times by wavefunction collapses \cite{carmichael93prl,open system,parkins95pra,mesoscopic QO},
we determine the probability of single-photon absorption as a function of relevant system parameters,
such as $\eta$ and the ratio of the spontaneous emission rates of the target emitter. In our Monte
Carlo wavefunction (MCWF) numerical simulations we have chosen a gaussian laser pulse with peak Rabi
frequency $\Omega_o$ which is of the order of $\Gamma_{31}^{(T)}$ and bandwidth which satisfies:
$\tau_{laser}^{-1}<\Gamma_{31}^{(T)}$. By keeping the quantity $\Omega_o/\Gamma_{31}^{(S)}$ small,
we assure that the generated single photon pulse is Gaussian and that it has a width
$\tau_{ph-pulse}^{-1}<\Gamma_{31}^{(T)}$. This assumption ensures
that the target atom is never saturated.

First we consider the case $\eta=1$: the only possible collapse
events in this case are due to the operators $\widehat{C}_1$ and
$\widehat{C}_3$. If $\Gamma_{31}^{(T)}\neq\Gamma_{32}^{(T)}$, there
is a finite probability for the two-emitters being projected into
either state $\left|1_S,1_T\right\rangle$ or
$\left|1_S,2_T\right\rangle$. However, if these two spontaneous
emission rates are equal, then the collapse events due to
$\widehat{C}_1$ never occur because of a destructive quantum
interference between the two paths leading to the state
$\left|1_S,1_T\right\rangle$.

To understand the origin of this interference effect, we first
consider the simpler case of weak coherent light incident on the
target atom. Following van Enk \cite{vanenk04pra} and
assuming $\eta = 1$, we find that the mean output field is
\begin{equation}
\left\langle \widehat{E}_{out}^{(+)}\right\rangle=\beta+\sqrt{\Gamma_{31}}\left\langle \sigma_{31}
\right\rangle = \beta - 2 \beta \frac{\Gamma_{31}^{(T)}}{\Gamma_{31}^{(T)} + \Gamma_{32}^{(T)}} .
\end{equation}
Here, we have evaluated $\left\langle \sigma_{31} \right\rangle$
using the Optical Bloch Equations for the 3-level target atom in
steady state. Clearly, $\left\langle
\widehat{E}_{out}^{(+)}\right\rangle = 0$ provided
$\Gamma_{31}^{(T)} = \Gamma_{32}^{(T)}$.

To analyze the single-photon
absorption we could alternatively use Optical Bloch Equations to evaluate $\left\langle
\widehat{C}_1^{\dagger}(t) \widehat{C}_1(t) \right\rangle =
\left\langle \widehat{E}_{out}^{(-)}
\widehat{E}_{out}^{(+)}\right\rangle$: in the limit
$\Gamma_{31}^{(T)} = \Gamma_{32}^{(T)}$ and $\Gamma_{31}^{-1} \ll
t \ll \Omega_o^{-1},\tau_{laser}$, we find analytically that
$\left\langle \widehat{C}_1^{\dagger}(t)\widehat{C}_1(t)
\right\rangle = 0$, implying perfect destructive interference.
Equivalently, the probability amplitude for detecting a photon
with energy $\omega_{31}^{(T)}$ that is emitted by the target has
the same magnitude but opposite sign as that of detecting a source
photon transmitted without scattering.

Given that for $\Gamma_{31}^{(T)} = \Gamma_{32}^{(T)}$ and $\eta=1$ the only possible
collapse event is the projection of the system into
$\left|a\right\rangle=\left|1_S,2_T\right\rangle$ and that all the other state occupancies
are decreased by a "no-collapse" event due to $H_{eff}$, the system
necessarily absorbs the incident photon. Since the target emitter
occupancy changes as a result of this process, we could also
consider the absorption process as "mapping of the incident photon
to an atomic excitation" \cite{fleischauer00prl}. Figure~2 shows the numerical calculations
based on an ensemble average of many quantum trajectories in the
limit $\eta = 1$: we observe that deterministic single-photon
absorption is robust against small variations around the value 1 of the
ratio $\Gamma_{32}^{(T)}/\Gamma_{31}^{(T)}$.

\begin{figure}[hb]
\includegraphics[width=0.35\textwidth]{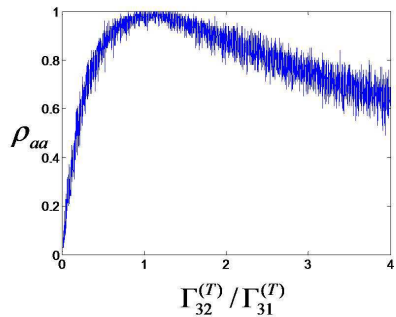}
\vspace{-0.3cm}
\caption{\small{$\Gamma_{32}^{(T)}/\Gamma_{31}^{(T)}$ dependance of the probability of
single photon absorption, defined
as the steady state expectation value of the desired state
($\left|a\right\rangle=\left|1_S,2_T\right\rangle$) population
($\rho_{aa}=\lim_{t\rightarrow\infty}\left\langle \sigma_{aa}(t)\right\rangle$). Here, $\Gamma_{31}^{(T)}=1$,
$\Gamma_{31}^{(S)}=10\Gamma_{31}^{(T)}$, $\gamma_{21}^{(T)}=0$,
$\Omega_L(t)=\Omega_o\exp{(-(t-t_0)^2/\tau^2)}$, $\Omega_0=1$, $\tau=10$, $t_0=20$.}}
\end{figure}

Next, we consider the dependence of the absorption probability on the dipole-emission overlap
parameter $\eta$: simulations depicted in Figure~3 confirm that the fidelity of single-photon
absorption is linearly proportional to $\eta$. 

\begin{figure}[ht]
\includegraphics[width=0.35\textwidth]{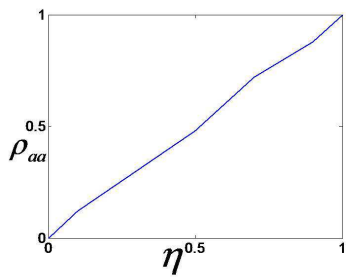}
\vspace{-0.5cm}
\caption{\small{Overlap parameter $\eta$ dependance of the probability of single photon
absorption. The parameters used in the simulation are the same as the ones used in Fig.2.}}
\end{figure}

To have a complete description of the dynamics, we would also need
a collapse operator $\widehat{C}_4
=\sqrt{\Gamma_{30}^{(S)}}\sigma_{03}^{(S)}$  describing light
scattering on the
$\left|3\right\rangle_S\rightarrow\left|0\right\rangle_S$
transition. We have set $\Gamma_{30}^{(S)}=0$ and thereby
neglected these collapse events since, up to this point, we have
been mainly interested in absorption by the target atom, of a
photon pulse that is generated by an {\bf ideal} source. If
$\Gamma_{30}^{(S)} \neq 0$, then the source atom will still act as
a unity-efficiency single-photon source, albeit with an
uncertainty (jitter) in the emission time of the single photon
pulse \cite{kiraz04pra}. This jitter however, has consequences for
single-photon absorption: we find in our simulations that choosing
$\Gamma_{30}^{(S)} = \Gamma_{31}^{(S)}$ leads to a $\sim 10\%$
decrease in the absorption probability of the target atom. This
decrease can be traced back to the fact that destructive
interference prohibiting collapse events described by
$\widehat{C}_1$, is no longer perfect.

One immediate application of near-deterministic single-photon absorption is in intra-conversion of a
single-photon polarization qubit to a localized (spin) qubit. We consider a generalization of the target
emitter energy level diagram for the case of polarization-spin conversion; we would in this case need
two optically excited target atom states $\left|3^{+}\right\rangle_T$ and
$\left|3^{-}\right\rangle_T$ which are coupled to the state $\left|1\right\rangle_T$ by a right and
left hand circularly polarized photon, respectively (the relevant target atom energy level diagram is
depicted in Figure~4). If these excited states decay into metastable states
$\left|2^{+}\right\rangle_T$ and $\left|2^{-}\right\rangle_T$ by emission of an identical photon,
respectively, then we would achieve the state transformation:
\begin{eqnarray}
\left|\Psi_{in}\right\rangle=|1\rangle_T\otimes(\kappa_1\hat{a}_{+}^{\dag}+\kappa_2\hat{a}_{-}^{\dag})\left|vac\right\rangle\longrightarrow\nonumber\\
\left|
\Psi_{final}\right\rangle=(\kappa_1|2^{+}\rangle_T+\kappa_2|2^{-}\rangle_T)\otimes\hat{b}^{\dag}\left|vac\right\rangle
\end{eqnarray}
upon deterministic absorption of the incident photon.  Here, $\hat{a}_{+}^{\dag}$,
$\hat{a}_{-}^{\dag}$ correspond to the photon creation operators, where the index {+} ({-}) labels
right (left) circular polarization and $\left|vac\right\rangle$ denotes the vacuum state of the
photon field. Similarly, $\hat{b}^{\dag}\left|vac\right\rangle$ denotes a photon emitted on the
$\left|3^{+}\right\rangle_T\rightarrow\left|2^{+}\right\rangle_T$ or
$\left|3^{-}\right\rangle_T\rightarrow\left|2^{-}\right\rangle_T$ transition of the target emitter.

In the case of $\eta < 1$, the intra-conversion will fail with a finite
probability; however, if the emitter has a recycling transition
typical for alkali atoms, then we can determine if the target is in ground
state $\left|1\right\rangle_T$ or in a subspace orthogonal to it, after the absorption
is completed. This would allow us to determine if the
single-photon absorption succeeded without compromising the faithful
intra-conversion, should we determine that it was successful.

\begin{figure}[ht]
\includegraphics[width=0.55\textwidth]{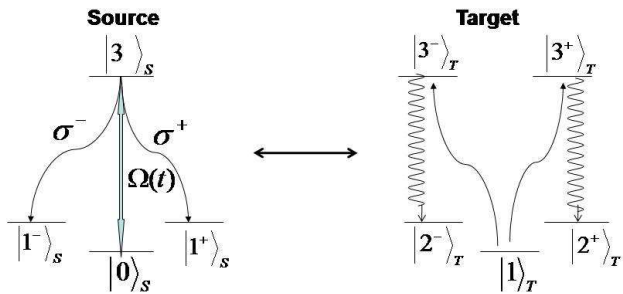}
\vspace{-2.5cm}
\caption{\small{Schematic of a free-space coupling scheme that achieves entanglement
of distant spin qubits. Following laser excitation, the source atom emits a single photon
whose polarization degree of freedom is entangled with the final source atom state.
Upon absorption of the single photon by the target, the two atoms are projected
onto a 'spin-entangled' state.}}
\end{figure}

Another potential application of the scheme analyzed in this work would be in the entanglement of
distant spin qubits with a high-success probability \cite{zhao07prl,duan01nature}. Until now, we have
considered an ideal source which generates a photon that can be collected with unity efficiency
($\eta_S=1$). This assumption has to be relaxed when we consider realistic spin entanglement. To this
end, we consider a source, which is initially in state $\left|0\right\rangle_S$ (Figure~4) and upon
linearly-polarized laser excitation generates a single right (left) hand circularly polarized photon
by Raman scattering into state $\left|1^{+}\right\rangle_S$ ($\left|1^{-}\right\rangle_S$). Clearly,
the end-state of the emission process is an emitter-photon entangled state
$\left|1^{+}\right\rangle_S \otimes \hat{a}_{+}^{\dag}\left|vac\right\rangle  +
\left|1^{-}\right\rangle_S \otimes \hat{a}_{-}^{\dag}\left|vac\right\rangle$. If the photon generated
by such a source emitter is absorbed by a target emitter as described earlier, then the end-result of
single-photon absorption is the generation of the entangled state, $\left|1^{+}\right\rangle_S
\otimes \left|2^{+}\right\rangle_T + \left|1^{-}\right\rangle_S \otimes \left|2^{-}\right\rangle_T$.
The success probability for generation of this state is $p\sim\eta^{2}$: for challenging but
realistic values of $\eta\sim0.3$ \cite{vanenk+kimble01pra,vanenk04pra}, this would yield an
entanglement generation probability of $\sim 10\%$, which compares favorably with the previously
proposed probabilistic long-distance entanglement schemes \cite{zhao07prl,duan01nature}. The ability
to carry out recycling transitions to determine whether or not the incoming single-photon is
absorbed, is crucial for this application \footnote{An advantage of the proposed scheme comes from the fact
that measurements based on atomic recycling transitions are more efficient than those based
on single photon detection.}.

Last but not least, we point out that our findings
could substantially simplify previous protocols for quantum state
transfer based on emitters embedded in cavities \cite{Ciraczoller97prl}. We consider a 
variant of the scheme of
Ref.~\cite{Ciraczoller97prl} and replace the coherent field
coupling ($\Omega_2(t)$) of the recipient (target) atom by
spontaneous emission $\Gamma_{re}$ on the $|r\rangle_2 -
|e\rangle_2$ transition \footnote{We use the notation of
Ref.~\cite{Ciraczoller97prl}~.}. We assume $\kappa \gg g$ and
adiabatically eliminate the cavity mode: in the Purcell limit $g^2
\gg \kappa \Gamma_{rg}$, we find that the cavity+atom system is
equivalent to an "effective single emitter" for which $\eta \sim
1$. If the Purcell enhancement is such that $\Gamma_{re} =
g^{2}/\kappa $, then the cavity+atom system can absorb the
incident single-photon with unity efficiency, without the need for
a coherent field that should have been synchronized with the
incoming single-photon pulse.

In conclusion, we have analysed the quantum state exchange between
light and matter at the level of single quanta. The formalism
presented here is general and can be applied to various quantum
systems such as atoms, QDs and molecules. The most important
challenge at this point is the enhancement of the overlap
parameter $\eta$; possible methods include change of the emission
profile of the dipole/atom by using dielectrics. Another
possibility involves a scheme proposed by Leuchs \textit{et al.}
\cite{leuchs00opt.comm.} to achieve tight focus of an incident
radially polarized field. For applications involving
intra-conversion of polarization and spin qubits, an additional
challenge is to achieve high $\eta$ for both polarizations
at the same time.

We wish to acknowledge many useful discussions with H. Tureci, S. van Enk, G. Leuchs, A. Sorensen and
V. Sandoghdar.

\end{document}